\documentclass[utf8,sort&compress]{frontiersFPHY}

\usepackage{color}
\usepackage{amsmath,amsfonts,amsthm,amssymb}
\usepackage{bm,bbm}
\usepackage[OT2,OT1]{fontenc}

\usepackage{url,hyperref,lineno,microtype}%,subcaption}
\usepackage[onehalfspacing]{setspace}

\newcommand{\be}{\begin{equation}}
\newcommand{\ee}{\end{equation}}
\newcommand{\ba}{\begin{eqnarray}}
\newcommand{\ea}{\end{eqnarray}}
\def\bs{\begin{subequations}}
\def\es{\end{subequations}}

\def\a{\alpha}

\def\s{\sigma}

\def\cL{\mathcal{L}}

\def\cob{\color{blue}}
\newcommand{\au}[2]{#1.~#2}
\newcommand{\book}[5]{\emph{#1}; #2: #3, #4, #5}
\newcommand{\books}[4]{\emph{#1}; #2: #3, #4}
\newcommand{\oarX}[1]{\href{http://arxiv.org/abs/#1}{{\cob arXiv:#1}}}
\newcommand{\arX}[1]{\href{http://arxiv.org/abs/#1}{{\cob arXiv:#1}}}
\newcommand{\doin}[6]{\href{http://dx.doi.org/#1}{{\cob {\it #2 #3} (#6) {\bf #4}:#5}}}
\newcommand{\doinn}[5]{\href{http://dx.doi.org/#1}{{\cob {\it #2} (#5) {\bf #3}:#4}}}
\newcommand{\doij}[5]{\href{http://dx.doi.org/#1}{{\cob {\it #2} (#5) {\bf #3}:#4}}}

\newcommand{\ndoinn}[5]{\href{#1}{{\cob {\it #2} (#5) {\bf #3}:#4}}}
\newcommand{\tia}[1]{#1.}

\newcounter{listcounter}

\def\keyFont{\fontsize{8}{11}\helveticabold }
\def\firstAuthorLast{Gianluca Calcagni}
\def\Authors{Gianluca Calcagni\,$^{*}$}
\def\Address{Instituto de Estructura de la Materia, CSIC, Serrano 121, 28006 Madrid, Spain}
\def\corrAuthor{Gianluca Calcagni}

\def\corrEmail{g.calcagni@csic.es}

\begin{document}
\twocolumn
\firstpage{1}

\title[Fundamental theory or data-driven models?]{Next step in gravity and cosmology: fundamental theory or data-driven models?}

\author[\firstAuthorLast]{\Authors} %This field will be automatically populated
\address{} %This field will be automatically populated
\correspondance{} %This field will be automatically populated

\extraAuth{}

\maketitle

%\begin{abstract}
%\tiny
{ \keyFont{Keywords: cosmology, inflation, gravitational waves, alternative theories of gravity, modified gravity, quantum gravity}} %All article types: you may provide up to 8 keywords; at least 5 are mandatory.
%\end{abstract}

\date{\small September 1, 2020}

%%%%%%%%%%%%%%%%%%%%%%%%%%%%%%%%%%%%%%%%%%
%%%%%%%%%%%%%%%%%%%%%%%%%%%%%%%%%%%%%%%%%%

The recent detection of gravitational waves (GWs) by LIGO-Virgo \cite{Abbott:2016blz} has opened a new window on the universe. Binary systems of compact objects such as black holes and neutron stars emit tiny ripples in spacetime when they inspiral and merge together. The ground-based interferometric system LIGO-Virgo has been measuring several such events and a new generation of experiments, both ground-based and space-borne, will be taking life in the next two decades, among which KAGRA \cite{Akutsu:2018axf} has just begun operations, LISA \cite{Audley:2017drz} is unfolding an ambitious science program, and projects such as the Einstein Telescope \cite{Maggiore:2019uih} and DECIGO \cite{Seto:2001qf,Kawamura:2011zz,Kawamura:2020pcg} are on the table. 
These interferometers will also open up the possibility to detect a stochastic GW background from the early universe, let it be from inflation, from cosmic phase transitions or from alternative scenarios. This stochastic background, given by the superposition of GWs of different amplitudes and phases coming from all directions in the sky, has not been observed so far \cite{TheLIGOScientific:2016wyq,Abbott:2017xzg} but future missions may be able to detect it \cite{Maggiore:2019uih,Kawamura:2020pcg,Bartolo:2016ami,Caprini:2019pxz}.

In parallel, GWs from astrophysical sources carry a wealth of precious information about their production (via a wave-form analysis) and propagation (via a modified dispersion relation and the luminosity distance) which, in turn, depend on the underlying gravitational theory. Thus, GW astronomy is an opportunity to test general relativity as well as theories beyond it \cite{Yunes:2016jcc,Belgacem:2019pkk}.

Testing Einstein gravity is a daunting task when it comes down to calculate the wave-form of an astrophysical event made of an inspiral, a merger and a ringdown phase involving two or more compact objects. However, can we use GWs also as a tool to check other gravitational theories? The answer would be in the affirmative if interferometers had the required sensitivity to discriminate between the predictions of different theories. This depends on the specific theory or model considered: some depart from general relativity so little that any exotic effect is completely negligible as far as GWs are concerned. Others carry larger modifications and may fall within the scope of future, or even near-future, experiments.

I employed purposefully vague expressions such as ``alternative scenarios'' or ``theories beyond general relativity'' to introduce the topic of this Grand Challenge. GWs entail many challenges, each of which is grand in its own, but here we focus on a question that, as a matter of fact, transcends the realm of GWs and relates to cosmology as a whole: Should we concentrate our effort on bottom-up (i.e., data-driven) models of gravity or on top-down models stemming from fundamental theories? Before giving you my personal take on the issue, let me first clarify terminology. By \emph{bottom-up} or data-driven models, I mean theoretical settings (the ``up'') created \emph{ad hoc} to explain one or more physical phenomena (the ``bottom'') and that, so far, have not been embedded in any fundamental theory. An example is the class of $f(R)$ models, which have been employed both as inflationary and as dark-energy scenarios \cite{DeFelice:2010aj,Clifton:2011jh,Capozziello:2011et}. On the other hand, \emph{top-down} or fundamental models are constructed in a fundamental theory of gravitation and/or particle physics (the ``top'') aiming to describe the elementary building blocks of geometry and matter and their interaction and, specifically, certain phenomena for which data are available (the ``down''). Examples are flux-compactification models of inflation in string cosmology (reviewed in, e.g., \cite{Baumann:2014nda,Calcagni:2017sdq}) and the bouncing cosmological scenario of loop quantum gravity \cite{Ashtekar:2011ni,Banerjee:2011qu,Gielen:2016dss}. In both cases, a cosmological model is derived from a theory of gravitation and matter. Models only inspired by, but not fully and rigorously embedded into, fundamental theories lie between the bottom-up and the top-down extrema. Examples are the first braneworld scenarios proposed when flux compactification was not yet under control \cite{Brax:2004xh,Maartens:2010ar}, the coeval cosmological tachyon scenarios (reviewed in \cite[section 13.7.2]{Calcagni:2017sdq}) or non-local cosmological scalars with operators resembling those found in the low-energy limit of string field theory \cite{Arefeva:2004qqr,Arefeva:2005mom}. Finally, the term \emph{phenomenology} can indistinctly refer either to top-down models closer than the mother theory to observed phenomena or to bottom-up models inspired by phenomena themselves. 

Paradoxically, top-down models may become bottom-up. Consider the rise and fall of Ho\v{r}ava--Lifshitz gravity. Born as a fundamental theory breaking Lorentz invariance \cite{Horava:2009uw}, it produced a rich cosmology studied in a flood of papers. When it was noticed that classical Lorentz symmetry breaking would amplify at the quantum level to the point of ruling out the theory \cite{Iengo:2009ix}, there were three types of reaction. A reduced group of theoreticians tried to modify the theory to avoid the problem; some lost interest in the theory; while all the rest, perhaps the majority, simply continued to do phenomenology with it, ignoring the issue.

The converse also happens and some models born as data-driven may turn out to be top-down when an embedding in a fundamental theory is discovered. An instance is Starobinsky gravity $\cL=R+\a R^2$ \cite{Starobinsky:1980te,Mukhanov:1981xt}. For years, it was only one among many exotic cosmological scenarios, until it came to prominence as the most favored model of primordial inflation \cite{Planck:2013jfk,Akrami:2018odb}. Soon afterwards, it was found that Starobinsky gravity could be recovered as the local limit of a unitary and renormalizable non-local theory of quantum gravity \cite{Briscese:2013lna,Koshelev:2016xqb}.

Starobinsky gravity is a special element of the class of $f(R)$ models. A limited number of other models may be derived from quantum gravity, for instance asymptotic safety \cite{Bonanno:2012jy,Hindmarsh:2012rc} or M-theory \cite{Nojiri:2003rz}, but their overwhelming majority cannot be embedded in a fundamental theory. The intuitive reason is that covariant gravitational interactions include Riemann--Riemann terms $R_{\mu\nu\rho\s} R^{\mu\nu\rho\s}$ and Ricci--Ricci tensor terms $R_{\mu\nu}R^{\mu\nu}$, which means that a theory made only of the Ricci scalar $R$ cannot reproduce consistently the quantum-corrected graviton propagator. 

To some, this is not a problem; to me, it is. The former could argue that their foremost objective is to fit data with a bottom-up model and, once successful, one can start to think about where such model could come from. Or even not! After all, general relativity does not ``come from'' anywhere in particular. It does have some special properties, such as being (up to topological terms) the most general four-dimensional theory of gravitation with second-order covariant equations of motion \cite{Lovelock:1971yv,Lovelock:1972vz}. However, this is hardly an explanation of its origin. It is assumed as a fundamental description of part of Nature, and this is it. Last but not least, the advocates of the primacy of data-driven models [such as $f(R)$ gravity] could also recall that progress in physics has often been made in a bottom-up fashion, from observations to theoretical models, and from them to full-fledged theories. In a sense, people often got an interest in rigorous theories when a first crude version of them showed to predict some phenomenon not understood until then. When Newtonian celestial mechanics proved inadequate to explain the precession of Mercury, Einstein's theory of gravitation could solve the problem. Just like general relativity was, in that case, a sort of extension of Newtonian gravity, so could $f(R)$ gravity be the natural extension of general relativity when it comes to explain the late-time acceleration of the universe.

It is undeniable that $f(R)$ gravity spurred a very fecund branch of research in cosmology which gave valuable or, at times, even invaluable insights into gravity and its possible extensions. On a theoretical level, this \emph{is} progress whose ramifications extended as far as string theory and other quantum gravities. However, pursuing a bottom-up model for its sake leads, in my opinion, to a dead end, even when the final outcome is positive and the model can explain data in a convincing way (i.e., per general consensus of the community). In that case, in fact, it is not guaranteed at all that it can fit into a bigger picture. Going back to my favorite scapegoat, a successful $f(R)$ would not be able to unify gravity and quantum mechanics into a theory of quantum gravity. It would not even be able to give a qualitative idea, or hints, of what such a theory should be. For example, the stability properties of quadratic gravity with only the $R^2$ term are very different from those of quadratic gravity with also Ricci--Ricci tensor and Riemann--Riemann terms \cite{Kawai:1998ab,Nunez:2004ts,Chiba:2005nz}. The first is not even a toy model of the latter. 

Ultimately, $f(R)$ models can reproduce dark energy \cite{Capozziello:2002rd}, but fail to explain it. Successful $f(R)$ models have inverse powers of the Ricci scalar and several parameters \cite{Hu:2007nk,Starobinsky:2007hu,Capozziello:2007eu,delaCruz-Dombriz:2015tye}, not fine tuned but introduced by hand. Reversing the logic, if a bottom-up model fails to fit data, one may tailor appropriate modifications to adjust reality better, and so on, and so on. As a whole, the resulting scenario escapes the Popperian notion of science and becomes unfalsifiable.

It would be unfair to keep bullying a class of models that, after all, have been long since replaced by other, perhaps more appealing candidates. So let me go after them, too. Horndeski models and DHOST theories are now a hot topic among cosmologists \cite{Langlois:2017mdk,Kase:2018aps,Kobayashi:2019hrl} and they may even have a say about GWs. For instance, their effect on the GW luminosity distance of standard sirens (sources of both GWs and photons) may be large enough to reach the sensitivity threshold of LISA \cite{Belgacem:2019pkk}. Also another phenomenological model, a non-local infrared modification of gravity, can produce interestingly large deviations from general relativity \cite{Belgacem:2020pdz}. So far, none of these models has been derived from or embedded into any fundamental theory. One could wave away this criticism with a two-pronged argument: 1) since bottom-up models carry a detectable signal, they can be very useful to study the science of future GW interferometers and to stimulate research in that direction; 2) despite oft-invoked theoretical ``evidence,'' quantum gravity is not a must and there might not be any such thing in Nature. At least, no observation so far forces this construct upon us.

We grant the latter point but, actually, the first is neither in favor nor against bottom-up models because it also applies to some quantum gravities. Reasonings loosely based on a curvature expansion of the action, or on the integration of short-scale degrees of freedom as in Wilsonian effective theory, or on continuum effective field theory\footnote{See \cite{Georgi:1994qn} for an explanation of the difference between Wilsonian and continuum effective field theory.}, concluding that no quantum-gravity effect can ever occur much above the Planck length scale, have been repeatedly refuted by models of inflation (reviewed in \cite[chapters 10, 11, 13]{Calcagni:2017sdq}) and GW propagation \cite{Yunes:2016jcc,Calcagni:2019ngc} where non-trivial non-perturbative mechanisms are in action. In quantum-gravity inflationary models, the accelerated expansion of the early universe boosts tiny corrections to cosmological scales leaving an imprint in the primordial scalar and tensor spectra; instances are several scenarios of string cosmology and loop quantum cosmology. Scenarios alternative to inflation also exist that fit data while having some characteristic features, such as string-gas cosmology \cite{Brandenberger:2015kga} and a new version of the ekpyrotic model \cite{Brandenberger:2020eyf}, although they are situated more in the middle of the bottom-up/top-down continuum. In some cases, the tensor spectrum is blue-tilted and it can be enhanced up to the sensitivity thresholds of GW interferometers, such as string-gas cosmology, the new ekpyrotic universe, and others \cite{CaKu}. Finally, although it is true that most quantum gravities predict corrections too tiny to be observed in late-time cosmological phenomena, in very few cases there is a chance (just a chance) that quantum geometry departs from the classical one enough to leave a cumulative effect on the GW luminosity distance, such as in group field theory, spin foams or loop quantum gravity \cite{Calcagni:2019ngc}.

Whether the community should focus mainly on data-driven models or on fundamental ones is a moot point: it is not an either-or choice and many models cannot be classified as sharply as I naively tried to do here. The actual Grand Challenge, perhaps, lies in making a bigger effort in justifying the available models from the perspective of fundamental interactions, regardless of their pedigree. And, of course, to have theoreticians and phenomenologists pay more and better attention to each other.

%%%%%%%%%%%%%%%%%%%%%%%%%%%%%%%%%%%%%%%%%%
%%%%%%%%%%%%%%%%%%%%%%%%%%%%%%%%%%%%%%%%%%

\section*{Acknowledgments}

\noindent{The author thanks S.\ Capozziello, S.\ Nesseris and E.\ Saridakis for useful comments, is supported by the I+D grant FIS2017-86497-C2-2-P of the Spanish Ministry of Science, Innovation and Universities, and acknowledges networking support by the COST Action CA18108.}

%%%%%%%%%%%%%%%%%%%%%%%%%%%%%%%%%%%%%%%%%%
%%%%%%%%%%%%%%%%%%%%%%%%%%%%%%%%%%%%%%%%%%

\end{document}